\documentclass[conference,a4paper,10pt]{IEEEtran}
\usepackage[keeplastbox]{flushend}
\usepackage[latin1]{inputenc}
\usepackage{times,amsmath}
\usepackage{amsfonts}
\usepackage{pstool}
\usepackage{subfigure}
\usepackage{multirow}
\usepackage{multicol}
\usepackage{enumerate}
\usepackage{graphicx}
\usepackage{MnSymbol}
\usepackage{stfloats}
\usepackage[table]{xcolor}
\usepackage[square, comma, sort&compress, numbers]{natbib}
\usepackage{nohyperref}
\usepackage{algorithm,algorithmic}
\usepackage{bigints}
\usepackage{amsmath}

\newcounter{tempEquationCounter}
\newcounter{thisEquationNumber}

\makeatletter
\newcommand{\vast}{\bBigg@{4}}
\newcommand{\Vast}{\bBigg@{5}}
\newcommand\numeq[1]%
  {\stackrel{\scriptscriptstyle(\mkern-1.5mu#1\mkern-1.5mu)}{=}}
\makeatother

\graphicspath{ {Figures/} }

\begin{document}


\title{On the Effective Capacity of an Underwater Acoustic Channel under Impersonation Attack}

\author{
\IEEEauthorblockN{Waqas\ Aman$^\ast$, Zeeshan\ Haider$^\ast$, S.\ Waqas\ H.\ Shah$^{\ast , \dagger}$, M.\ Mahboob\ Ur\ Rahman$^\ast$, Octavia A. Dobre$\ddagger$\\
$^\ast$Electrical engineering department, Information Technology University, Lahore 54000, Pakistan \\
$^{\dagger}$Computer Laboratory, University of Cambridge, Cambridge CB3 0FD, U.K. \\
$^\ddagger$Department of Electrical and Computer engineering, Memorial University, St.  John's, NL A1B 3X5, Canada \\
$^\ast$\{waqas.aman, waqas.haider, mahboob.rahman\}@itu.edu.pk, $^\ddagger$odobre@mun.ca
}
}

\maketitle


\maketitle

\begin{abstract}

This paper investigates the impact of authentication on effective capacity (EC) of an underwater acoustic (UWA) channel. Specifically, the UWA channel is under impersonation attack by a malicious node (Eve) present in the close vicinity of the legitimate node pair (Alice and Bob); Eve tries to inject its malicious data into the system by making Bob believe that she is indeed Alice. To thwart the impersonation attack by Eve, Bob utilizes the distance of the transmit node as the feature/fingerprint to carry out feature-based authentication at the physical layer. Due to authentication at Bob, due to lack of channel knowledge at the transmit node (Alice or Eve), and due to the threshold-based decoding error model, the relevant dynamics of the considered system could be modelled by a Markov chain (MC). Thus, we compute the state-transition probabilities of the MC, and the moment generating function for the service process corresponding to each state. This enables us to derive a closed-form expression of the EC in terms of authentication parameters. Furthermore, we compute the optimal transmission rate (at Alice) through gradient-descent (GD) technique and artificial neural network (ANN) method. Simulation results show that the EC decreases under severe authentication constraints (i.e., more false alarms and more transmissions by Eve). Simulation results also reveal that the (optimal transmission rate) performance  of the ANN technique is quite close to that of the GD method.

\end{abstract}


\section{Introduction}
\label{sec:intro}
Underwater acoustic sensor networks (UWASN) are utilized by a multitude of applications, e.g., resource finding, marine-life exploration, marine pollution monitoring, and security of oil rigs \cite{Akyildiz:AdNetworks:2005}, \cite{Felemban:IJDSN:2015}. As of today, a wide range of (theoretical and hands-on) research problems related to UWASN have been reported in the literature, e.g., channel capacity, the acoustic modem design, routing protocols, full-duplex, source localization, to name a few \cite{Milica:RS:2012,Waqas:AINA:2016,Dobre:OCEANS:2018,Dobre:WCL:2019}. 

This work studies the two-pronged challenge of {\it secure and reliable communication over an underwater acoustic (UWA) channel}. The security challenge arises because the UWA channel--being a broadcast medium--is prone to various kinds of attacks by adversaries. On the other hand, the reliability challenge implies that the data sent on the UWA channel is delay-sensitive, and thus, quality-of-service (QoS) constrained. 

Security front first: There are only a handful of works that discuss the security requirements, classification of various active and passive attacks, and potential solutions for the UWASN. As one would expect, most of the security solutions are crypto-based, while very few works discuss physical-layer security solutions (to complement the crypto-based security solutions at the higher layer) \cite{Security:IWC:2011}. This paper considers impersonation attack on a UWASN and thwarts it by utilizing distance as a feature for doing feature-based authentication at the physical layer. Note that various (device-based or medium-based) fingerprints/features have been reported in the literature on physical layer security, e.g., received signal strength \cite{Trappe:TPDS:2013}, channel impulse response \cite{Ammar:VTC:2017}, channel frequency response \cite{Xiao:TWC:2008}, path loss, carrier frequency offset \cite{Rahman:Globecom:2014}, distance, angle-of-arrival, position \cite{Aman:Access:2018}, non-reciprocal hardware \cite{Mahboob:VTCS:2017}, and non-linearity of power amplifiers \cite{Dobre:TIFS:2016}.

Next, the reliability front: One way to quantify the QoS offered by a UWA link is by computing the effective capacity (EC)---maximum throughput of the channel under QoS constraints. More formally, the EC is the maximum sustainable {\it constant} arrival rate at a transmitter (queue) in the face of a randomly time-varying (channel) service, under QoS constraints \cite{Wu:TWC:2003}. The EC tool has witnessed its application to a diverse set of problems for QoS performance analysis, e.g., cognitive radio channels \cite{Gursoy:TWC:2010}, \cite{Anwar:TVT:2016}, systems with various degrees of channel knowledge at the transmitter \cite{gross2012scheduling}, two-hop systems \cite{Gursoy:TIT:2013}, \cite{Lateef:TC:2009}, correlated fading channels \cite{Soret:TWC:2010}, and device-to-device communication \cite{WShah:WCL:2019}. 

The work closest to the scope of this paper is \cite{Henrik:Globcom:2017} and its extension \cite{Henrik:JSAC:2019}, whereby the impact of authentication on the delay performance of mission-critical, machine-type communication networks has been reported. Specifically, \cite{Henrik:Globcom:2017} and \cite{Henrik:JSAC:2019} assume a closed and protected environment (whereby Eve's transmissions do not reach Bob) and utilize the tool of stochastic network calculus to quantify the dependence of delay violation probability on false alarms (inherited in the physical layer authentication mechanism). Contrary to \cite{Henrik:Globcom:2017} and \cite{Henrik:JSAC:2019}, this paper considers a UWA channel and  an open environment for communication (whereby Bob can receive messages from Eve as well) and utlizes the EC tool. In short, to the best of the authors' knowledge, {\it the impact of authentication constraints on the EC of a UWA channel has not been studied in the literature so far}.

{\bf Contributions}: The contribution of this paper is two-fold. 1) We investigate the impact of authentication on the EC of a UWA multipath channel. To this end, we provide a closed-form expression of the EC in terms of authentication parameters. 2) We formulate the optimal transmission rate problem as an optimization program, and solve it via gradient-descent and artificial neural network methods.

{\bf Outline}: Section II introduces the system model, problem statement and the UWA multipath channel model. Section III describes the distance-based authentication mechanism. Section IV presents the EC analysis. Section V computes the optimal transmission rate using the gradient-descent and artificial neural network methods. Section VI provides some numerical results. Section VII concludes the paper.

\section{System Model and Channel Model}
\label{sec:sys-model}

\subsection{System Model and Problem Statement}
\begin{figure}[ht]
\begin{center}
	\includegraphics[width=3.0in]{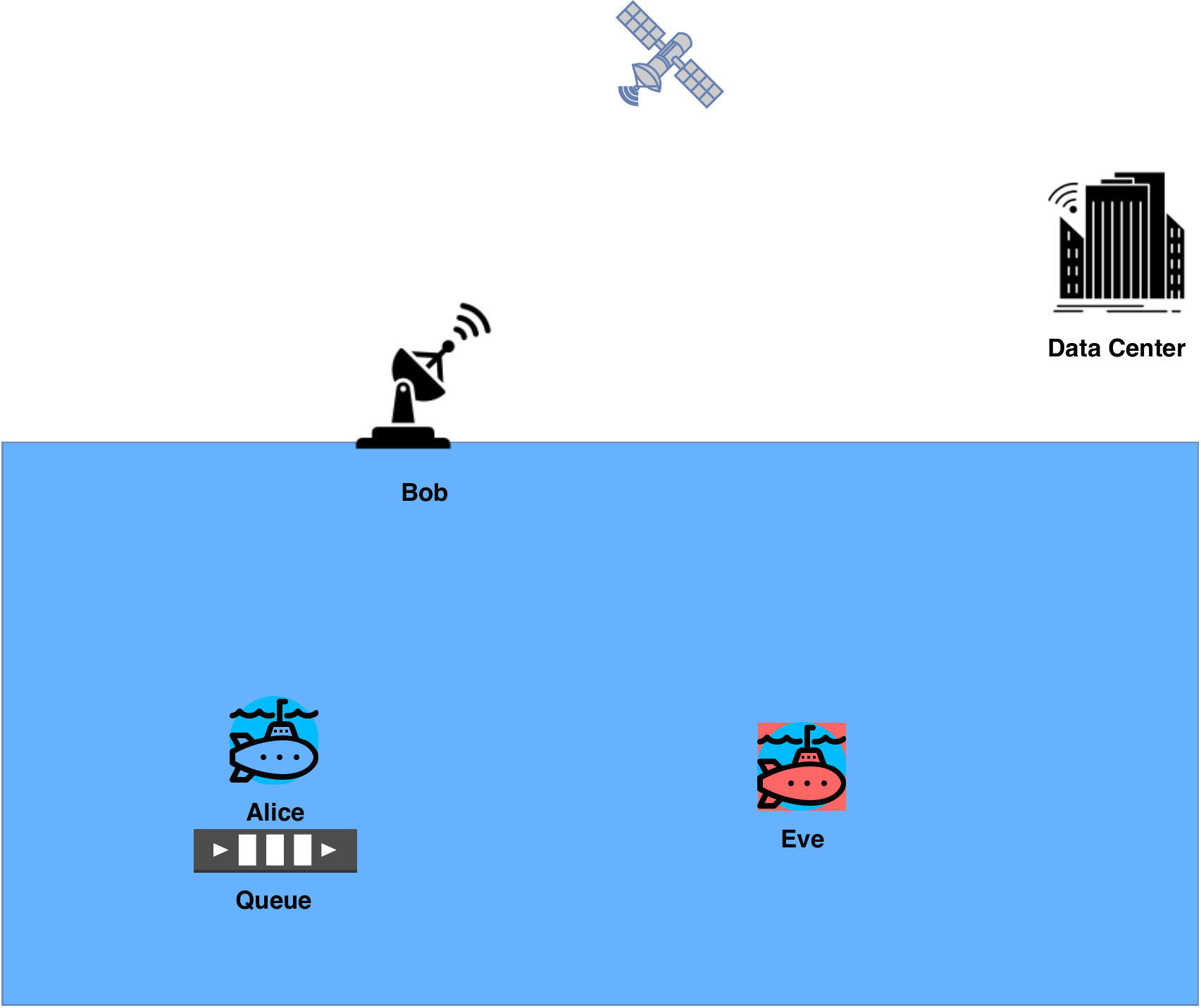}
\caption{System model: The UWA link between the legitimate node pair (Alice and Bob) is QoS-constrained, and is under impersonation attack by a nearby adversary Eve.}
\label{fig:sm}
\end{center}
\end{figure}

We consider a scenario whereby a sensor node (Alice) deployed underwater reports its sensing data to a buoy node (Bob) on the water surface, via a time-slotted UWA channel (see Fig. \ref{fig:sm}). For the UWA link between the legitimate node pair (Alice and Bob), we study the two-pronged challenge of {\it secure and reliable communication}. More precisely, the security threat studied in this paper is impersonation attack, while the QoS of the underlying shared UWA channel is assessed and quantified via the EC framework. 

As shown in Fig. \ref{fig:sm}, the impersonation attack is led by Eve, which is a malicious node present underwater in the close vicinity of Alice. That is, Eve is an active adversary who tries to inject malicious data into the system (i.e., into the data center through Bob). More precisely, Eve---being a clever impersonator, and not a mere jammer---transmits during the slots left unused by Alice\footnote{Eve can accomplish this through spectrum sensing technique.} in order to deceive Bob that she is indeed Alice. This necessitates that Bob authenticates each and every packet it receives from the shared UWA channel. To this end, Bob measures the {\it distance} of the UWA channel occupant (Alice or Eve) from itself and utilizes it as the feature/fingerprint for binary hypothesis testing in order to carry out feature-based authentication at the physical layer. On the reliability front, the EC framework allows us to compute the maximum constant arrival rate at the Alice's queue in the face of fading multipath UWA channel, given a statistical QoS constraint.

Note that the authentication process results in occasional rejection of Alice's (delay-sensitive) data by Bob due to false alarms. This in turn requires re-transmission of that data by Alice in the upcoming slot, and hence, affects the QoS provided by the considered UWA link. {\it Thus, studying the interplay between authentication and EC is of utmost importance, and this is precisely the agenda of this paper.}

\subsection{Channel Model}

We consider a wideband, time-slotted UWA channel with $T$ seconds long time-slots. Furthermore, the transmit nodes (Alice and Eve) utilize orthogonal frequency division multiplexing (OFDM) scheme with $N$ sub-carriers, while $T_g$ is the guard interval between two consecutive  OFDM symbols. Thus, the effective slot length is $T_s=T+T_g$.

For the multipath UWA channel, we adopt the widely-acclaimed, statistical model from \cite{Aval:Ucomms:2014}. Let $H(f)$ represent the transfer function/channel frequency response (CFR) of the underlying UWA channel. The values of $H(f)$ at sub-carrier frequencies $f^{(i)} = f_0 + i \Delta f$ (where $i=0,...,N-1$) are denoted by $H^{(i)}=H(f^{(i)})$ and assumed to be constant over a subband of width $\Delta f = 1/T$.   
Furthermore, the channel gain $H^{(i)}$ seen by sub-carrier $i$ is composed of $L$ paths as follows \cite{Aval:Ucomms:2014}:
\begin{equation}
H^{(i)}=\frac{1}{\sqrt{\mathcal{A}}}\sum_{l=1}^L h_l e^{-j2\pi f^{(i)} \xi_l}, 
\end{equation}
where $\mathcal{A}$ is the total attenuation (due to spreading and absorption), while $h_l$ and $\xi_l$ are the path gain and the path delay of the $l^{th}$ path, respectively.
Furthermore, $H^{(i)} \sim CN(\sum_{l=1}^Lc_l\mathbb{E}\{h_l\},\sigma_L^2\sum_{l=1}^L\vert c_l\vert )$ \cite{Aval:Ucomms:2014}. Here, $c_l= \frac{1}{\sqrt{\mathcal{A}}}e^{-j2\pi f^{(i)} \xi_l}$, $\sigma_L = \sigma_l \ \ \forall l$, $CN$ stands for complex normal distribution, and $\mathbb{E}(.)$ is the expectation operator.
\section{Impersonation Attack Detection}

This section succinctly describes the {\it distance-based} authentication framework implemented by Bob to thwart the impersonation attack by Eve. Specifically, we provide here a brief sketch of the binary hypothesis test constructed by Bob and the associated error probabilities. 

\subsection{Distance-based Authentication}
During each slot, Bob makes a noisy measurement of the distance $d_{n}$ of the channel occupant (Alice or Eve) from itself.\footnote{\cite{Aman:Access:2018} describes a time-of-arrival/round-trip time-based method for distance estimation in a UWA channel.} With this, Bob implements the distance-based authentication as the following binary hypothesis test (BHT):
\begin{equation}
	\label{eq:H0H1}
	 \begin{cases} H_0: & d_{n}=d_A + e \; \; \; \text{(Declare Alice)} \\
                   H_1: & d_{n}=d_E + e \; \; \; \text{(Declare Eve)}, \end{cases}
\end{equation}
where $d_A$ ($d_E$) represents the true distance of Alice (Eve) from Bob, and $e$ is the estimation error. Under least-squares estimation framework, $e\sim N (0,\sigma^2)$.

Let $\mathcal{T} = d_{n}-d_{A}$. Further, let $\tau = |\mathcal{T}|$ be the test statistic. Then the BHT in (\ref{eq:H0H1}) could be (re-)formulated as: $\tau \underset{H_0}{\overset{H_1}{\gtrless}} \epsilon$, where $\epsilon$ is a threshold. Essentially, Bob compares the noisy distance measurement $d_{n}$ with the pre-stored ground truth $d_A$. Thereafter, if the test statistic $\tau$ is less than the threshold $\epsilon$, then $H_0$ occurs; otherwise, $H_1$ occurs. 

Note from (\ref{eq:H0H1}) that $d_{n}|H_0\sim N(d_{A}, \sigma^2)$, and $d_{n}|H_1\sim N(d_{E}, \sigma^2)$. Then, $\mathcal{T} \mid H_0 \sim N(0,\sigma^2)$ and $\mathcal{T} \mid H_1 \sim N(d_{E}-d_{A},\sigma^2)$. With this, we are ready to compute the two error probabilities associated with the BHT: false alarms and missed detections. The probability of false alarm (wrongly rejecting Alice's data) is:
\begin{equation}
P_{fa} = P(H_1|H_0) = P(\tau >\epsilon|H_0) = P(|\mathcal{T}| >\epsilon|H_0) = Q(\frac{\epsilon}{\sigma}),
\end{equation}
where $Q(x)=\frac{1}{\sqrt{2\pi}} \int_x^\infty  e^{-\frac{t^2}{2}} dt$ is the complementary cumulative distribution function (CDF) of ${N}(0,1)$. Similarly, the probability of missed detection (success rate of Eve) is:
\begin{equation}
P_{md} = P(H_0|H_1) = P(\tau <\epsilon|H_1) = P(|\mathcal{T}| <\epsilon|H_1) = 1 - Q(\frac{\epsilon-m}{\sigma}),
\end{equation}
where $m=d_{E}-d_{A}$. Note that $P_{md}$ is a random variable (RV) since we do not know $d_{E}$. Thus, we find the expected value of $P_{md}$ by assuming that $d_{E}\sim U(a,b)$, i.e., Eve is neither too close, nor too far away from Bob:
\begin{equation}
\bar{P}_{md}=\mathbb{E}\{P_{md}\}=\frac{1}{(b-a)} \int_a^b P_{md}dd_{E}.
\end{equation} \\
Next, to compute the threshold $\epsilon$, we utilize the Neyman-Pearson (NP) method which states that one cannot minimize the two errors $P_{fa}, P_{md}$ simultaneously. The NP method, however, guarantees to minimize $P_{md}$ once a maximum-tolerable value of $P_{fa}$ is pre-specified. This allows us to systematically compute the threshold $\epsilon$ as follows:
\begin{equation}
\epsilon = \sigma Q^{-1}(P_{fa}).
\end{equation}
Finally, the Kullback-Leibler divergence (KLD) is a measure of how reliable the distance measurements are (and thus, the BHT). The KLD $D(p(\tau|H_1)||p(\tau|H_0))$ is given as: $D = \int_{-\infty}^{\infty} p(\tau|H_1) \log(\frac{p(\tau|H_1)}{p(\tau|H_0)}) d\tau = -\frac{m}{\sigma^2}$.

\section{Effective Capacity Analysis}
\label{sec:EH}

\subsection{Definition}

The EC is defined as the log of moment-generating function (MGF) of the cumulative service process $S^{(i)}(t)$ in the limit (for sub-carrier $i$) \cite{Wu:TWC:2003}:
\begin{equation}
\label{eq:ECStandard}
\text{EC}_i = -\frac{\Lambda^{(i)} (-\theta)}{\theta}=-\lim_{t \to \infty} \frac{1}{\theta t} \ln \{ \mathbb{E} (e^{-\theta S^{(i)}(t)}) \} \; [\text{bits/slot}],
\end{equation}
where $S^{(i)}(t)=\sum_{k=1}^t s^{(i)}(k)$, with $s^{(i)}(k)$ being the channel service (i.e., number of bits delivered) on sub-carrier $i$ during slot $k$. $\theta$ is the so-called QoS exponent.

The EC could be thought as the maximum constant arrival rate that can be supported by a randomly time-varying channel service process, while also satisfying a statistical QoS requirement specified by the QoS exponent $\theta$. 
Moreover, since the average arrival rate at Alice's queue is equal to the average departure rate when the queue is in steady-state \cite{Chang:Infocom:1995}, EC can also be seen as the maximum throughput in the presence of QoS constraints. One could also see that $\theta \to 0$ implies delay-tolerant communication, while $\theta \to \infty$ implies delay-limited communication. 

\subsection{Implications of Lack of Channel Knowledge at Alice}

We assume that the channel state information at the transmitter (CSIT) is not available at Alice (and Eve). In other words, the system is not equipped with a feedback channel that Bob could utilize to share the measured CSI to the channel occupant. Let $r_A^{(i)}$ represent the Alice's data rate on sub-carrier $i$. Then, due to lack of CSIT, Alice transmits data to Bob at a constant rate on all sub-carriers, i.e., $r_A^{(i)}=r_{A,c}$ bits/sec for $i=1,...,N$. Furthermore, the lack of CSIT prompts Alice to distribute its power budget $P_A$ equally among the $N$ sub-carriers, i.e., $P_A^{(i)}=\frac{P_A}{N}$.\footnote{Due to lack of CSIT, it is reasonable to assume that Eve also transmits at a fixed rate on all sub-carriers, i.e., $r_E^{(i)}=r_{E,c}$ bits/sec for $i=1,...,N$. Furthermore, the lack of CSIT also prompts Eve to do equal power allocation among the $N$ sub-carriers, i.e., $P_E^{(i)}=\frac{P_E}{N}$.} 

As for the communication link between the channel occupant and Bob, a threshold-based (ON/OFF) error model is considered. That is, assuming that Alice is the channel occupant, if the fixed rate $r_A^{(i)}$ is less than the instant channel capacity $C_A^{(i)}(k)$ during the slot $k$, then the link conveys $r_A^{(i)}$ bits/sec and is said to be in ON condition. On the other hand, if $r_A^{(i)}$ is greater than $C_A^{(i)}(k)$, the link conveys 0 bits/sec and is said to be in OFF condition.\footnote{When the link is in OFF condition, the bits sent by the channel occupant need to be re-transmitted (e.g., using the automatic repeat request mechanism) during the next slot.}

In short, due to authentication at Bob, due to lack of CSIT at the channel occupant, and due to the threshold-based error model, the relevant dynamics of the considered system could be modelled by a Markov chain.

\subsection{Markov Chain Representation of the UWA Channel}
\label{subsec:MC}

\begin{center}
\begin{table}
    \caption{Markov chain representation of the dynamics of the UWA channel faced by the sub-carrier $i$.}
    \begin{tabular}{ | c | p{4cm} | p{3cm} |}
    \hline
    State & Description & Notation \\ \hline
    $1$ & Bob correctly detects Alice and the link is ON ($s_i(k)=r_A^{(i)}T_s$) & $H_0|H_0$ \& $r_A^{(i)}<C_A^{(i)}(k)$ \\ \hline
    $2$ & Bob correctly detects Alice and the link is OFF ($s_i(k)=0$) & $H_0|H_0$ \& $r_A^{(i)}>C_A^{(i)}(k)$ \\ \hline
    $3$ & Bob correctly detects Eve and the link is ON ($s_i(k)=0$) & $H_1|H_1$ \& $r_E^{(i)}<C_E^{(i)}(k)$ \\ \hline
    $4$ & Bob correctly detects Eve and the link is OFF ($s_i(k)=0$) & $H_1|H_1$ \& $r_E^{(i)}>C_E^{(i)}(k)$ \\ \hline
    $5$ & Bob wrongly detects Alice as Eve and the link is ON ($s_i(k)=0$) & $H_1|H_0$ \& $r_A^{(i)}<C_A^{(i)}(k)$ \\ \hline
    $6$ & Bob wrongly detects Alice as Eve and the link is OFF ($s_i(k)=0$) & $H_1|H_0$ \& $r_A^{(i)}>C_A^{(i)}(k)$ \\ \hline
    $7$ & Bob wrongly detects Eve as Alice and the link is ON    ($s_i(k)=r_E^{(i)}T_s$) & $H_0|H_1$ \& $r_E^{(i)}<C_E^{(i)}(k)$ \\ \hline
    $8$ & Bob wrongly detects Eve as Alice and the link is OFF ($s_i(k)=0$) & $H_0|H_1$ \& $r_E^{(i)}>C_E^{(i)}(k)$ \\ 
    \hline
    \end{tabular}
\end{table}
\end{center}

Table I describes the Markov chain representation (with eight states) of the UWA channel faced by the sub-carrier $i$. 
For the states $1-8$ defined in Table I, let $\textbf{P}^{(i)}$ represent the transition probability matrix with
$[\mathbf{P}]^{(i)}_{u,v}=p_{u,v}^{(i)}$ being the transition probability from state $u$ (at slot $k-1$) to state $v$ (at slot $k$). Note that the state of the link changes after a slot duration, i.e., $T_s$ (due to block fading). Next, we compute the state transition probabilities, starting with $p_{1,1}$:\footnote{We drop the sub-carrier index $i$ for simplicity of notation.}
\begin{equation}
\label{eq:p11}
\begin{split}
p_{1,1} = P \{ &\tau|H_0(k)<\epsilon \; \& \; r_A^{(i)}<C_A^{(i)}(k) \; \big| \\
\; &\tau|H_0(k-1)<\epsilon \; \& \; r_A^{(i)}<C_A^{(i)}(k-1) \}.
\end{split}
\end{equation}

The computation of $p_{1,1}$ in (\ref{eq:p11}) could be simplified as follows: 
\begin{equation}
\begin{split}
p_{1,1} \numeq{a} &P \{ \tau|H_0(k)<\epsilon \; \big| \; \tau|H_0(k-1)<\epsilon \}. \\
&P \{ r_A^{(i)}<C_A^{(i)}(k) \big| r_A^{(i)}<C_A^{(i)}(k-1) \} \\
\numeq{b} &P \{ \tau|H_0(k)<\epsilon \} P \{ r_A^{(i)}<C_A^{(i)}(k) \}, \\
\end{split}
\end{equation}
where the equality {\it (a)} follows from the fact that the fading process $\{C_A^{(i)}\}_k$ is independent of the authentication process $\{\tau\}_k$, and {\it (b)} follows from the fact that each of the two stochastic processes is memoryless, i.e., $C_A^{(i)}(k)|C_A^{(i)}(k-1)=C_A^{(i)}(k)$ and $\tau(k)|\tau(k-1)=\tau(k)$. 

Let $\gamma_A^{(i)}$ represent the signal-to-noise ratio (SNR) of the link seen by the sub-carrier $i$ when Alice transmits.\footnote{Since both stochastic processes are memoryless, we drop the time index $k$ for notational simplicity.} Specifically, $\gamma_A^{(i)} = \frac{P_A^{(i)}\vert H_A^{(i)}\vert^2}{\sigma_n^2}$ with $\sigma_n^2$ as the variance of the circularly-symmetric complex Gaussian noise. Next, since $C_A^{(i)}=\Delta f \log_2(1+\gamma_A^{(i)})$, computing $P \{ r_A^{(i)}<C_A^{(i)} \}$ is equivalent to computing $P\{\gamma_A^{(i)}>2^{\frac{r_A^{(i)}}{\Delta f}}-1\}$, which is simply the complementary CDF of $\gamma_A^{(i)}$ evaluated at $2^{\frac{r_A^{(i)}}{\Delta f}}-1$. 

{\it Proposition 4.1:} $\gamma_A^{(i)}\sim \chi_{2}^2 (\lambda_A^{(i)})$. That is, $\gamma_A^{(i)}$ is  non-central chi-squared distributed with two degrees-of-freedom and the non-centrality parameter $\lambda_A^{(i)} = \frac{2P_A^{(i)}\mathcal{A}}{\sigma_n^2 L}\mid \sum_{l=1}^L c_{l,A} E\{h_{l,A} \} \mid^2$. 

{\it Proof:} See Appendix A.

Due to Proposition 4.1, $P \{ r_A^{(i)}<C_A^{(i)}(k) \}=(Q_1(\sqrt{\lambda_A^{(i)}},\sqrt{2^{\frac{r_A^{(i)}}{\Delta f}}-1}))$, where $Q_{x}(.,.)$ is the Marcum Q-function with $\lambda_A^{(i)} = \frac{2P_A^{(i)}\mathcal{A}}{\sigma_n^2 L}\mid \sum_{l=1}^L c_{l,A} E\{h_{l,A} \} \mid^2$ and degree $x$. Next, $P \{ \tau|H_0(k)<\epsilon \} = P(H_0|H_0)=\pi (A) (1-P_{fa})$, where $\pi (A)$ is the prior probability of Alice. Therefore: 
\begin{equation}
p_{u,1} = p_1=\pi (A) (1-P_{fa})(Q_1(\sqrt{a},\sqrt{b})),
\end{equation}
where $a=\lambda_A^{(i)}$,$b=2^{\frac{r_A^{(i)}}{\Delta f}}-1$. Note that the state-transition probability $p_{1,1}$ does not depend on the original state (which is $1$). In general, $p_{u,1} = p_1$ for any state of origin $u$. Furthermore, due to Proposition 4.1, $\gamma_E^{(i)}\sim \chi_{2}^2 (\lambda_E^{(i)})$ with $\lambda_E^{(i)} = \frac{2P_E^{(i)}\mathcal{A}}{\sigma_n^2 L}\mid \sum_{l=1}^L c_{l,E} E\{h_{l,E} \} \mid^2$.
Therefore:
\begin{equation}
\label{eq:p2p3p4}
\begin{split}
& p_{u,2} = p_2 = \pi (A)(1-P_{fa})(1-Q_1(\sqrt{a},\sqrt{b})) \\
& p_{u,3} = p_3 = \pi (E)(1-P_{md})(Q_1(\sqrt{c},\sqrt{d}))  \\
& p_{u,4} = p_4 = \pi (E)(1-P_{md})(1-Q_1(\sqrt{c},\sqrt{d})) \\
& p_{u,5} = p_5 = \pi (A)(P_{fa})(Q_1(\sqrt{a},\sqrt{b})) \\
& p_{u,6} = p_6 = \pi (A)(P_{fa})(1-Q_1(\sqrt{a},\sqrt{b})) \\
& p_{u,7} = p_7 = \pi (E)(P_{md})(Q_1(\sqrt{c},\sqrt{d})) \\
& p_{u,8} = p_8 =\pi (E) (P_{md})(1-Q_1(\sqrt{c},\sqrt{d})), \nonumber
\end{split}
\end{equation}
where $c=\lambda_E^{(i)}$,$d=2^{\frac{r_E^{(i)}}{\Delta f}}-1$. $\pi (E)$ is the prior probability of Eve, $\pi (E)=1-\pi (A)$ (i.e., Eve utilizes those slots which are idle). With this, each row of $\mathbf{P}^{(i)}$ becomes: $[p_1, p_2, p_3, p_4, p_5,p_6,p_7,p_8]$. Note that $\mathbf{P}^{(i)}$ has rank 1 due to identical rows, and is a stochastic matrix (since the sum along each row is equal to $1$).\\

\subsection{Effective Capacity of the UWA Channel}

With entries of $\mathbf{P}^{(i)}$ computed, we utilize the following result to calculate the EC of the sub-carrier $i$ \cite{Chang:TNCS:2012}:
\begin{equation}
  \frac{\Lambda^{(i)} (\theta)}{\theta}=\frac{1}{\theta} \ln (sp(\mathbf{\Phi}^{(i)}(\theta)\mathbf{P}^{(i)})).
\end{equation}
The above result states that for a Markov service process $S^{(i)}(t)$ with its dynamics modelled by $\mathbf{P}^{(i)}$, the MGF is given as $sp(\mathbf{\Phi}^{(i)}(\theta)\mathbf{P}^{(i)})$, where $sp(.)$ represents the spectral radius of a matrix and $\mathbf{\Phi}^{(i)}(\theta)$ is a diagonal matrix which contains the MGFs of the service process in the eight states as its diagonal elements. 

Note from Table I that $s^{(i)}=r_A^{(i)}T_s$ bits for state $1$, $s^{(i)}=r_E^{(i)}T_s$ bits for state $7$, and $s^{(i)}=0$ bits for the remaining states. Accordingly, the MGF of state $1$ is $e^{\theta r_A^{(i)}T_s}$, MGF of state $7$ is $e^{\theta r_E^{(i)}T_s}$, while the MGF for each of the remaining states is $1$. Thus, $\mathbf{\Phi}^{(i)}(\theta)=\text{diag}([e^{\theta r_A^{(i)}T_s},1,1,1,1,1,e^{\theta r_E^{(i)}T_s},1])$. Next, since $\mathbf{\Phi}^{(i)}(\theta)\mathbf{P}^{(i)}$ is also a matrix of unit-rank, finding its spectral radius is equivalent to finding its trace. Thus, the EC/throughput (bits/sec) of sub-carrier $i$ under statistical QoS and security constraints is:
\begin{equation}
\label{eq:EC}
\text{EC}_i = \frac{-1}{\theta T_s} \big[ \ln ( p_1e^{\theta r_A^{(i)} T_s}+ p_2+ p_3+ p_4+ p_5+ p_6+p_7e^{\theta r_E^{(i)} T_s}+ p_8) \big].
\end{equation}

Finally, the net EC/throughput for the considered UWA channel (i.e., OFDM with $N$ sub-carriers) is given as \cite{Ge:TVT:2014}:
\begin{equation}
\label{eq:ECOFDM} 
\text{EC}_{tot} = \sum_{i=1}^N \text{EC}_{i}.
\end{equation}

\section{Optimal Transmission Rate of Alice}
\label{sec:rate}

Equation (\ref{eq:EC}) reveals that Alice could further optimize its transmission rate (for sub-carrier $i$) as follows: $r_{A}^{(i)*}=\arg \max_{r_A^{(i)}>0} \text{EC}_{i}$. That is:
\begin{equation}
\label{eq:ECiopt}
    r_{A}^{(i)*}=\arg \max_{r_A^{(i)}>0} \frac{-1}{\theta T_s} \big[ \ln ( p_1e^{\theta r_A^{(i)} T_s}+ \sum_{w=2}^6 p_w +p_7e^{\theta r_E^{(i)} T_s}+ p_8) \big].
\end{equation}
Equivalently, we have:
\begin{equation}
\label{eq:opt}
    r_{A}^{(i)*}=\arg \min_{r_A^{(i)}>0} \big[ p_1e^{\theta r_A^{(i)} T_s}+ \sum_{w=2}^6 p_w +p_7e^{\theta r_E^{(i)} T_s}+ p_8 \big].
\end{equation}
Table 1 reveals that Eve is active during states $3,4,7,8$; therefore, the transition probabilities $p_3,p_4,p_7,p_8$ are irrelevant when optimizing  (\ref{eq:opt}) with respect to (w.r.t.) ${r_A^{(i)}}$. Discarding the irrelevant terms and simplifying the remaining terms, we get:
\begin{align}
\label{eq:opt2}
  r_{A}^{(i)*}=\arg \min_{r_A^{(i)}>0} P_{c,A}(e^{\theta r_A^{(i)} T_s} - 1)Q_1(\sqrt{a},\sqrt{b}),
\end{align}
where $P_{c,A} = \pi(A)(1-P_{fa})$ is the probability of correctly detecting Alice. Let $C_f=P_{c,A}(e^{\theta r_A^{(i)} T_s} - 1)Q_1(\sqrt{a},\sqrt{b})$.

\subsection{Gradient-descent based Approach}

One can verify that the cost function $C_f$ in  (\ref{eq:opt2}) is a convex function \cite{Yu:ITSP:2011}. Thus, taking its derivative w.r.t. ${r_A^{(i)}}$ and using the product rule, chain rule and the result of \cite{Yu:ITSP:2011}, we get: 
\begin{align}
\label{eq:deropt}
  \frac{\partial C_f}{\partial r_A^{(i)}} = P_{c,A}Q_1^`(a,b)(e^{\theta T_s r_A^{(i)}}-1)+P_{c,A}Q_1(a,b)(e^{\theta T_s r_A^{(i)}}\theta T_s),
\end{align}
where $Q_1^`(a,b)=\frac{\partial Q_1(a,b)}{\partial r_A^{(i)}}=-\frac{I_0(ab)e^{-\frac{(a^2+b^2)}{2}} \ln(2)2^{\frac{r_A^{(i)}}{\Delta f}}}{2\Delta f }$ and $I_0(.)$ is the zero-order modified Bessel function.
Now, setting $\frac{\partial C_f}{\partial r_A^{(i)}}$ equal to zero, we get:
\begin{align}
\label{eq:dCdr}
\frac{Q_1(a,b)}{I_0(ab)e^{\frac{-(a^2+b^2)}{2}}2^{\frac{r_A^{(i)}}{\Delta f}}}= \frac{ \ln(2) (e^{\theta T_s r_A^{(i)}}-1)}{2\Delta f \theta T_s e^{\theta T_s r_A^{(i)}}}.
\end{align}
Deriving a closed-form solution for $r_A^{(i)}$ from  (\ref{eq:dCdr}) is quite involved (because $Q_1(a,b)$ and $I_0(ab)$ both contain infinite number of terms). Luckily, we have the gradient in the hand; therefore, we compute the optimal rate $r_{A}^{(i)*}$ through iterative gradient-descent (GD) method. The control-law for the GD method is given as: 
 
\begin{align}
r_A^{(i)}(m)=r_A^{(i)}(m-1)-\alpha\Delta \big|_{r_A^{(i)}(m-1)},
\end{align}
where $m$ is the iteration number, $\alpha$ is the step-size and $\Delta$ is the gradient of the cost function $C_f$ (see (\ref{eq:deropt})).

\subsection{Artificial Neural Network based Approach}
The GD method iteratively solves the arrival rate optimization program, while the number of iterations depends on the initialization of the variables and the step size. In this subsection, we solve the transmit rate optimization problem as a regression problem by leveraging the artificial neural network (ANN) approach. The motivation behind proposing an ANN-based solution for optimal rate prediction is to realize a fast and computationally less-expensive algorithm as compared to the iterative GD solution. 

\begin{figure}[ht]
\begin{center}
	\includegraphics[width=3.0in,height=1.8in]{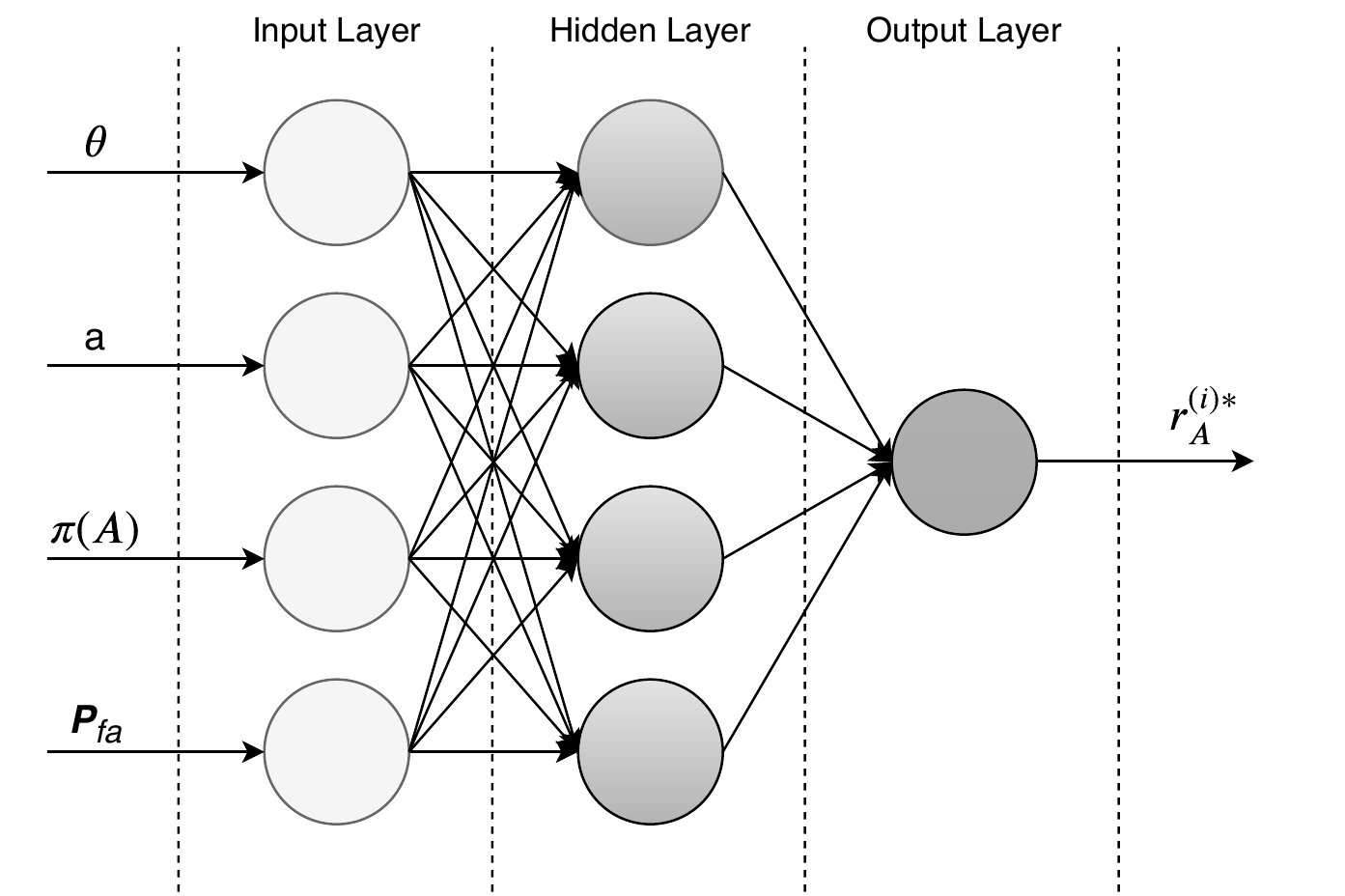}
\caption{The proposed ANN architecture. }
\label{fig:ann}
\end{center}
\end{figure}

Fig. \ref{fig:ann} shows the proposed ANN architecture with $3$ layers: an input layer with $4$ neurons, a hidden layer with $4$ neurons, and an output layer with one neuron. Thus, the number of parameters that are to be learned by this ANN is $25$ (hidden layer has $20$ parameters, while the output layer has $5$ parameters). The set of input features consists of $\theta$, $a$, $P_{fa}$, and $\pi(A)$. 

In order to train the ANN, we generated a dataset with input feature vectors and passed it to the iterative GD method which returned the optimal rate labels. We used the rectified linear unit (ReLu) as an activation function at the hidden layer (which enforces the constraint on the arrival rate, i.e., it should be positive), and the loss function as mean square error (MSE) at the output layer. ReLu can be expressed as $f(z)=max(0,z)$, where $z$ is the input. The loss function is given as: $\frac{\sum_{i=1}^{R}{(Y_{actual}^{i}-Y_{predicted}^{i})^2}}{R}$, where $Y_{actual}^{i}$ is the label of the training data, $Y_{predicted}^{i}$ is the output of ANN for a specific test input, and $R$ is the size of the dataset.
\section{Numerical Results}
\label{sec:results}

All simulations (Matlab-based and Python-based for ANN) are performed using an Intel Core i$7$-$4770$ octa-core processor with the $8$ GB memory. We consider $T=50$ ms long slots, and use OFDM scheme with $N=256$ sub-carriers with sub-carrier spacing $\Delta f=20$ Hz and guard interval $T_g =16$ ms \cite{Aval:Ucomms:2014}. Each of the CFR tap gain $\vert H_i\vert$ is generated from the Rice distribution with shape parameter $K=1$. Finally, we set $\sigma^2=1$ and $\sigma_n^2=1$.

\begin{figure}[ht]
\begin{center}
	\includegraphics[width=3.6in,height=2.4in]{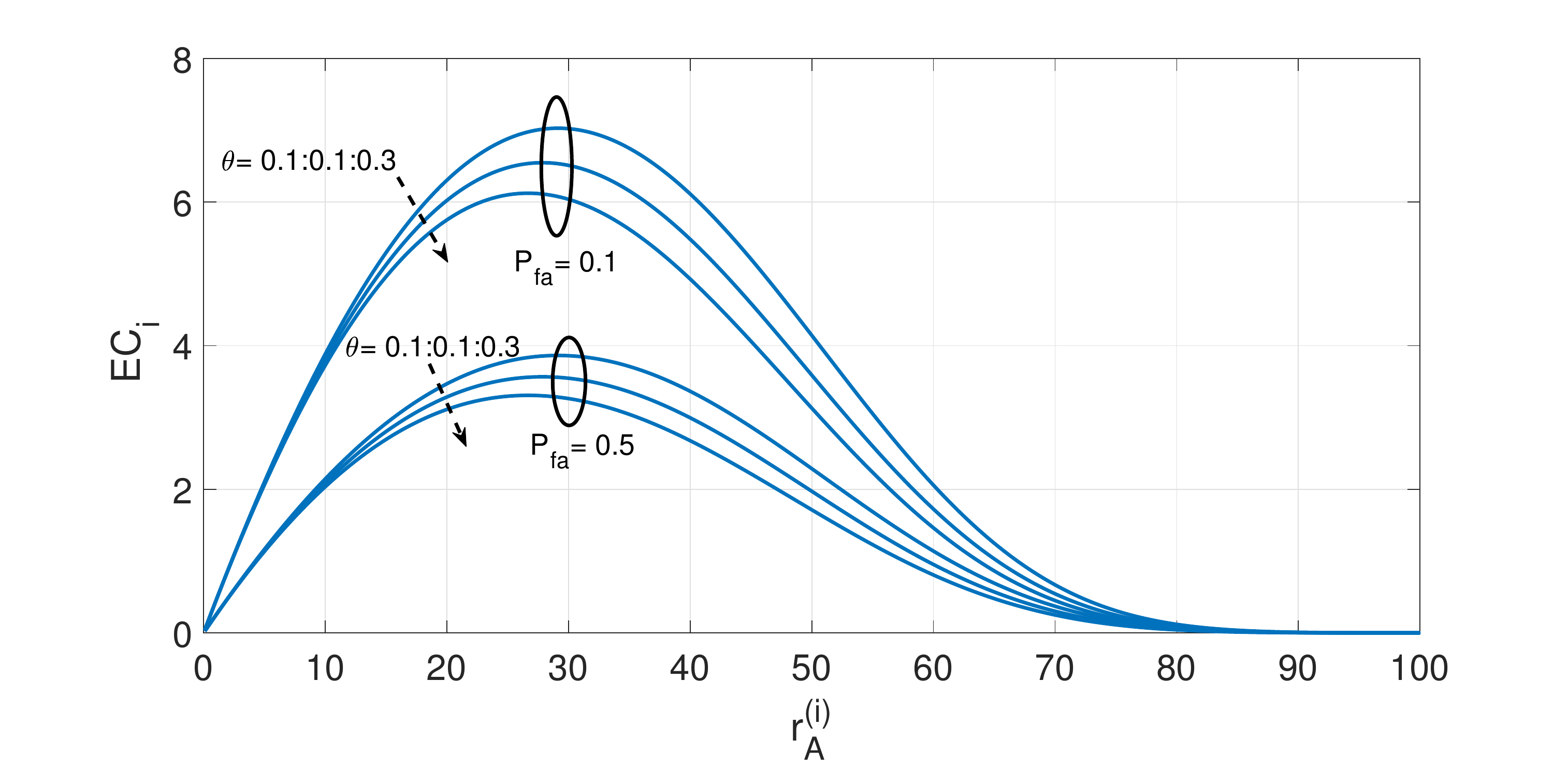}
\caption{$\text{EC}_i$ is a concave function of $r_A^{(i)}$. To obtain this plot, an exhaustive search over $r_A^{(i)}$ was performed in (\ref{eq:ECiopt}) under equal priors, i.e., $\pi(A)=\pi(E)$. }
\label{fig:ecvsr}
\end{center}
\end{figure}

Fig. \ref{fig:ecvsr} verifies that  $\text{EC}_i$ is indeed a concave function of $r_A^{(i)}$, i.e., an optimal transmission rate $r_A^{(i)*}$ does exist. Additionally, Fig. \ref{fig:ecvsr} reveals that the $\text{EC}_i$ decreases with the increase in $P_{fa}$ (as expected). We further notice that the optimal transmission rate $r_A^{(i)*}$ increases slightly as $\theta$ (the QoS constraint) increases.

\begin{figure}[ht]
\begin{center}
	\includegraphics[width=3.8in,height=2.6in]{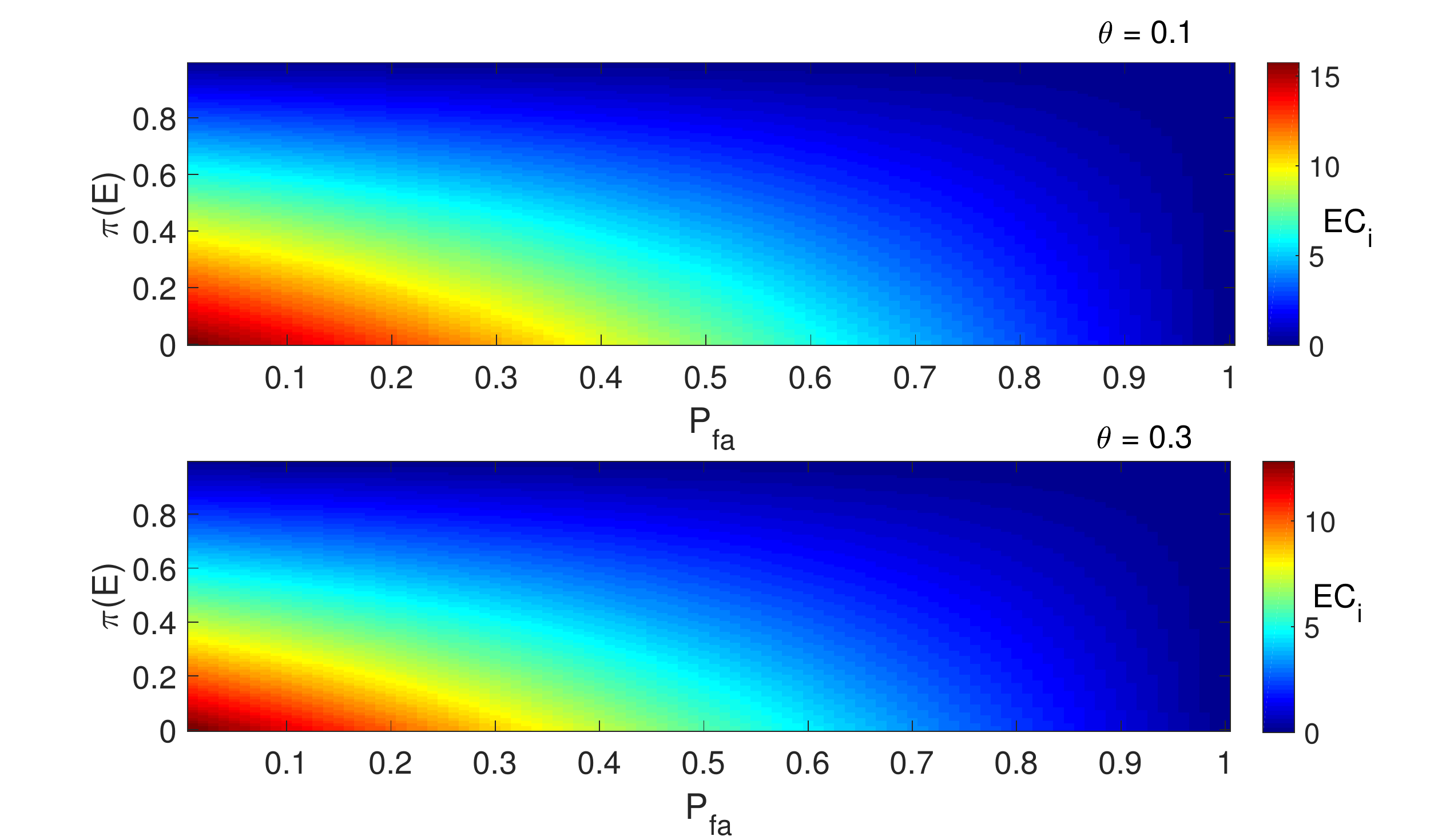}
\caption{Impact of authentication parameters ($P_{fa}$ and $\pi(E)$) on $\text{EC}_i$.  }
\label{fig:ec3d}
\end{center}
\end{figure}

Fig. \ref{fig:ec3d} studies the impact of authentication on the EC. Specifically, Fig. \ref{fig:ec3d} demonstrates that the EC decreases as the authentication constraints become more severe (i.e., with increase in either the probability of false alarm or the probability of transmission of Eve), and vice versa. Additionally, an increase in $\theta$ leads to a reduction in the EC.  

\textbf{ANN Setup}: For training data, we generate $90000\times 4 $ input samples along with $90000\times 1$ corresponding output labels. Specifically, the input dataset is generated by varying each of $\theta$, $\pi(A)$, $P_{fa}$ using a step size of $0.1$, while keeping the remaining features fixed. As for the non-centrality parameter $a$ of the SNR $\gamma_A^{(k)}$, we generate $100$ random samples of $a$ for each value of the remaining feature set. 
The ANN is trained at a learning rate of $0.001$ for $150$ epochs. This learning rate is reduced by the factor of $10$ after every $25$ epochs. Training data is further divided into training set and validation set for every epoch. Specifically, during each epoch, $80\%$ training data is used for training the ANN and the remaining $20\%$ is used as validation set. The MSE loss between the actual output and the predicted output is used to steer the ANN in the right direction.   

\begin{figure}[ht]
\begin{center}
	\includegraphics[width=3.6in,height=2.4in]{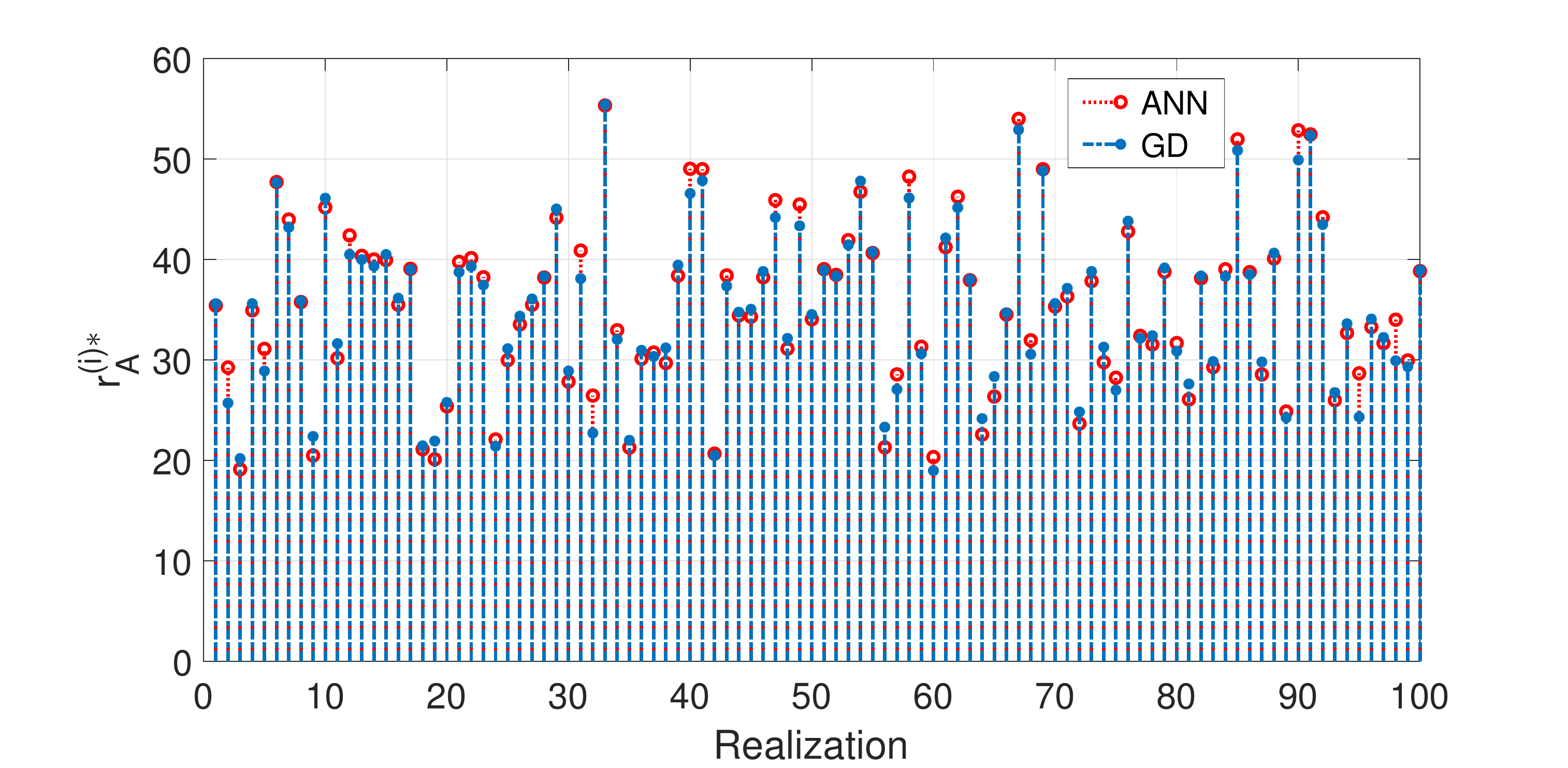}
\caption{Comparison of the GD method and the ANN method.  }
\label{fig:gdann}
\end{center}
\end{figure}

Fig. \ref{fig:gdann} plots the optimal transmission rate predicted by the ANN when a test input (basically, $100$ random samples/realizations of the input features) is applied, and compares it against the optimal rate returned by the GD method. Fig. \ref{fig:gdann} clearly shows that the ANN method performs very close to the GD method. The recorded MSE was around $1.8$ for different test data sets of the same size. 

Last but not the least, we also determined the simulation time (using the tic-toc command) to compute the optimal transmission rate $r_A^{(i)}$, for the GD method and the ANN method. Specifically, as we changed $N$ in the range $[32, 64, 128, 256]$ sub-carriers, the simulation times were recorded to be $[2.79, 5.70, 11.08, 22.21]$ seconds for the GD method, and $[0.02, 0.03, 0.06, 0.13]$ seconds for the ANN method. Thus, the ANN method is roughly two orders of magnitude faster than the GD method.\footnote{Note that the time-complexity of the ANN method is considered for the test data only. The reason of omitting the training time is that we train the ANN only once, while for the GD method, the iteration-based mechanism remains unchanged whenever a new input (feature) sample is provided.}
\section{Conclusion}
\label{sec:conclusion}
This work studied the trade-off between authentication and EC for a UWA channel that was under impersonation attack by a malicious node Eve. Specifically, a closed-form expression of the EC was derived as a function of authentication parameters. Furthermore, the optimal transmission rate (at Alice) was computed using the GD method and the ANN method. Simulation results showed that more transmissions by Eve and more false alarms (i.e., more severe authentication constraints) reduce the EC---the QoS provided by the UWA channel, and vice versa.

Future work will study the scenarios when the transmit nodes (Alice and Eve) have various degrees of channel knowledge (e.g., full, partial, and statistical).

\section*{Acknowledgement}
This work was partially supported by Natural Sciences and Engineering Research Council of Canada, through its Discovery program.
\appendices
\section{Proof of Proposition 4.1}

The channel gain at sub-carrier $i$ is $H^{(i)} \sim CN(\sum_{l=1}^L c_l \mathbb{E}\{h_l\},\sigma_L^2\sum_{l=1}^L\vert c_l\vert )$, where $c_l= \frac{1}{\sqrt{\mathcal{A}}}e^{-j2\pi f_i\xi_l}$ and $\sigma_L = \sigma_l \ \ \forall l$. Next, $\vert H^{(i)}\vert$ is Ricean distributed with shape parameter $K=\frac{\vert\sum_{l=1}^L c_l \mathbb{E}\{h_l\}\vert^2}{\sigma_L^2\sum_{l=1}^L\vert c_l\vert }=\frac{\vert\sum_{l=1}^L c_l \mathbb{E}\{h_l\}\vert^2}{\sigma_L^2 \frac{L}{\mathcal{A}} }$. Equivalently, $\vert H^{(i)}\vert \sim \text{Rice}(\sqrt{\frac{2\mathcal{A}}{L}}\mid \sum_{l=1}^L c_l \mathbb{E}\{h_l\}\mid , \sigma_L)$. Assuming unit variance for the path's distribution (i.e., $\sigma_l^2 = 1$), $\vert H^{(i)}\vert^2$ is distributed as non-central chi-squared with two degrees-of-freedom and non-centrality parameter $\left( \sqrt{\frac{2\mathcal{A}}{L}} \mid \sum_{l=1}^L c_l \mathbb{E}\{h_l\}\mid \right)^2$. Finally, $\gamma^i \sim \chi_{2}^2\left( \frac{2P^{(i)} \mathcal{A}}{\sigma_n^2L}\mid \sum_{l=1}^L c_l \mathbb{E}\{h_l\} \mid^2\right)$. Adopting the notation for Alice, $\gamma_A^i \sim \chi_{2}^2\left( \frac{2P_A^{(i)} \mathcal{A}}{\sigma_n^2 L}\mid \sum_{l=1}^L c_{l,A} \mathbb{E}\{h_{l,A}\} \mid^2\right)$.

\footnotesize{
\bibliographystyle{IEEEtran}
\bibliography{references}
}

\vfill\break

\end{document}